\documentstyle[prb,aps,epsfig]{revtex}
\begin{document}

\title{Sea-Boson Theory of Landau Fermi Liquids \\ 
Luttinger Liquids and Wigner Crystals}
\author{Girish S. Setlur \\ The Institute of Mathematical Sciences \\ 
 Taramani, Chennai 600113}

\maketitle

\begin{abstract}
 It is shown how Luttinger liquids may be studied using sea-bosons.
 The main advantage of the sea-boson method is its 
 ability to provide information about short-wavelength physics in 
 addition to the asymptotics and is naturally generalisable
 to more than one dimension. 
 In this article, we solve the Luttinger model
 and the Calogero-Sutherland model, the latter in the weak-coupling limit.
 The anomalous exponent we obtain in the former case is
 identical to the one obtained by Mattis and Lieb.
 We also apply this method to solve the two-dimensional analog 
 of the Luttinger model and show that the system is a Landau Fermi liquid.
 Then we solve the model of spinless fermions in one-dimension with long-range
 (gauge) interactions and map the Wigner crystal phase of the system.
\end{abstract}

\section{Introduction}

 In this article, we show how Luttinger liquids may
 be obtained from sea-bosons.
 This rectifies some serious errors of judgement of previous works
 \cite{Setlur1}. In addition, we are able to solve for the one-particle
 properties of various systems highlighted in the abstract.
 Specifically, we are able to write down a
 closed formula for the momentum distribution
 of the Wigner crystal mentioned in the abstract that does not involve
 arbitrarily chosen momentum cutoffs. Thus our formalism is able to provide
 information about short-wavelength physics as well as the asymptotics.
 The subject of fermions in one-dimension is vast and we shall only provide
 some representative references. Many of the more important references
 on this subject have already been included in our earlier work\cite{Setlur1}.
 Some others are by Haldane\cite{Hald}, Schulz, Cuniberti
 and Pieri\cite{Schulz1} and Schulz\cite{Schulz2}.

\section{ The Sea-Boson Method }

 In this article, we lay down the prescription for computing the momentum
 distribution of Fermi systems using the sea-boson method.
 We shall be quite cryptic at times since by now the terminology is well-known
 \cite{Setlur1} $ \mbox{  }^{,} $ \cite{Setlur2}. 
 In our earlier work\cite{Setlur2}, we
 made it quite clear that the sea-displacement annihilation
 operator has to be defined as follows.
 \begin{equation}
 A_{ {\bf{k}} }({\bf{q}}) = \frac{
 n_{F}({\bf{k}}-{\bf{q}}/2)(1-n_{F}({\bf{k}}+{\bf{q}}/2))
 }{\sqrt{n_{ {\bf{k}}-{\bf{q}}/2 }}}c^{\dagger}_{
 {\bf{k}}-{\bf{q}}/2 }c_{ {\bf{k}} + {\bf{q}}/2 }
 \label{AKQ}
 \end{equation}
 This definition is ambiguous due to the square root of the
 number operator in the denominator, but leaving that aside (see
 the earlier work\cite{Setlur2} for more details), it is
 clear that we may use this to rewrite the
 number operator\cite{Setlur2} as follows
 (where $ n_{F}({\bf{k}})  = \theta(k_{F}-|{\bf{k}}|) $ is the momentum 
 distribution of the free Fermi gas),
\begin{equation}
n_{ {\bf{k}} } = n_{F}({\bf{k}}) \mbox{           }n^{<}_{
{\bf{k}} } + (1-n_{F}({\bf{k}}))\mbox{           }n^{>}_{ {\bf{k}}
} \label{NUMSUGG}
\end{equation}
Here,
\begin{equation}
n^{<}_{ {\bf{k}} } = {\bf{ {\hat{1}} }} - \sum_{ {\bf{q}} \neq 0
}A^{\dagger}_{ {\bf{k}} + {\bf{q}}/2 }({\bf{q}})A_{ {\bf{k}} +
{\bf{q}}/2 }({\bf{q}})
\label{NLESS}
\end{equation}
\begin{equation}
n^{>}_{ {\bf{k}} } = \sum_{ {\bf{q}} \neq 0 }A^{\dagger}_{
{\bf{k}} - {\bf{q}}/2 }({\bf{q}})A_{ {\bf{k}} - {\bf{q}}/2
}({\bf{q}})
\label{NMORE}
\end{equation}
and $ {\bf{ {\hat{1}} }} = {\hat{N}}/N^{0} $.
 Here  $ N^{0} = \sum_{ {\bf{k}} } n_{F}({\bf{k}}) $
 and $  {\hat{N}} = \sum_{ {\bf{k}} }c^{\dagger}_{ {\bf{k}} }c_{ {\bf{k}} } $.
 From
Eq.(~\ref{NUMSUGG}) it is clear that in general we should expect a
discontinuity in the momentum distribution at
 $ |{\bf{k}}| = k_{F} $ since in general,
 $ \langle n^{<}_{ {\bf{k}}_{F} } \rangle \neq  \langle n^{>}_{ {\bf{k}}_{F} }
 \rangle $. This means that we may
  regard Eq.(~\ref{NUMSUGG}) as a proof of the Luttinger theorem. 
  This is true no matter how exactly or approximately
  we have solved our equations.
  It seems therefore that Landau Fermi liquids are generic.
  But there is a way in which this theorem can be violated.
  The only way in which
 Luttinger liquids can arise for arbitrarily weak repulsion is if
 the sums over $ {\bf{q}} $ in Eq.(~\ref{NMORE}) and
 Eq.(~\ref{NLESS}) diverge in such a way as to compensate for the
 individual terms in the sum being vanishingly small.
 Indeed, Eq.(~\ref{NUMSUGG}) suggests a natural classification of
 spinless one component Fermi liquids.
 We may consider the functions $ \langle n^{>}_{ {\bf{k}} } \rangle $
 and $ \langle n^{<}_{ {\bf{k}} } \rangle $ to be well defined
 and nonzero in general,
 (possibly with infinite slope at $ |{\bf{k}}| = k_{F} $)
 for all $ {\bf{k}} $ after factoring
 out the discontinuous $ n_{F}({\bf{k}}) $ and $ 1-n_{F}({\bf{k}}) $.

\noindent {\bf{Defn}} : If we have,

(i) $ \langle n^{<}_{ {\bf{k}}_{F} } \rangle > 
 \langle n^{>}_{ {\bf{k}}_{F} } \rangle $
: Landau Fermi liquid.

(ii)  $ \langle n^{<}_{ {\bf{k}}_{F} } \rangle  =
 \langle n^{>}_{ {\bf{k}}_{F} } \rangle $
: Residueless Fermi system.

(iii) $ \langle n^{<}_{ {\bf{k}}_{F} } \rangle < \langle n^{>}_{
{\bf{k}}_{F} } \rangle $ : Inverse Landau Fermi liquid.

\noindent It seems that the third possibility has not been investigated at
all, and with good reason.
 Later in this article, we argue that there is no room for such a
possibility, the most exotic ones are Luttinger liquids (or Wigner crystals).

 The operator in Eq.(~\ref{AKQ}) is not an exact boson but we may treat it
 as such and impose canonical boson commutation rules
 (this time we include spin for the sake of generality) :
$ [A_{ {\bf{k}} \sigma }({\bf{q}}\sigma^{'}), 
 A^{\dagger}_{ {\bf{k}} \sigma }({\bf{q}}\sigma^{'})] = 
 n_{F}({\bf{k}}-{\bf{q}}/2)(1-n_{F}({\bf{k}}+{\bf{q}}/2)) $.
 However now it appears that we must pay a price for this luxury. 
 The reason is quite subtle. Treating this object as an exact boson 
 and using the formulas for the momentum distribution
 written down in the main text of our first work
 leads to logarithimic divergences. In order to tame these divergences
 we have to adopt the following prescription.
\begin{equation}
\langle n_{ {\bf{k}}\sigma } \rangle = n_{F}( {\bf{k}} )\mbox{
}\langle n^{<}_{ {\bf{k}}\sigma } \rangle
 + (1-n_{F}({\bf{k}}))\mbox{           }
\langle n^{>}_{ {\bf{k}}\sigma } \rangle
 \label{NKSIG}
\end{equation}
\begin{equation}
\langle n^{<}_{ {\bf{k}}\sigma } \rangle =  \frac{1}{2} 
\left[ 1 + \frac{1}{ exp[2 S^{0}_{B}({\bf{k}}\sigma)] } \right] 
\label{F1}
\end{equation}
\begin{equation}
\langle n^{>}_{ {\bf{k}}\sigma } \rangle = \frac{1}{2} 
\left[ 1 - \frac{1}{ exp[2 S^{0}_{A}({\bf{k}}\sigma)] } \right] 
\label{F2}
\end{equation}
By definition, we have,
\begin{equation}
S^{0}_{B}({\bf{k}}\sigma) =
\sum_{ {\bf{q}} \sigma_{1} }
\langle G_{0} | a^{\dagger}_{ {\bf{k}}+{\bf{q}}/2 \sigma }({\bf{q}}\sigma_{1})
a_{ {\bf{k}}+{\bf{q}}/2 \sigma }({\bf{q}}\sigma_{1}) | G_{0} \rangle
(1-n_{F}({\bf{k}}+{\bf{q}}))
\label{S0B}
\end{equation}
\begin{equation}
S^{0}_{A}({\bf{k}}\sigma) =
\sum_{ {\bf{q}} \sigma_{1} }
\langle G_{0} | a^{\dagger}_{ {\bf{k}}-{\bf{q}}/2 \sigma_{1} }({\bf{q}}\sigma)
a_{ {\bf{k}}-{\bf{q}}/2 \sigma_{1} }({\bf{q}}\sigma) | G_{0} \rangle
n_{F}({\bf{k}}-{\bf{q}})
\label{S0A}
\end{equation}
We define, $A_{ {\bf{k}}\sigma }({\bf{q}}\sigma^{'}) = n_{F}({\bf{k}}-{\bf{q}}/2)
(1-n_{F}({\bf{k}}+{\bf{q}}/2))a_{ {\bf{k}}\sigma }({\bf{q}}\sigma^{'}) $.
Here $ |G_{0} \rangle $ is the ground state of the full hamiltonian with the
the sea-displacement operators being treated as canonical bosons with
the commutation rule
 $ [a_{ {\bf{k}}\sigma }({\bf{q}}\sigma^{'}),a^{\dagger}_{ {\bf{k}}\sigma }({\bf{q}}\sigma^{'})] = 1 $ ;
 and all other commutators involving any two of these operators are zero.
 Also it involves ignoring the presence of
 the square root making the Fermi bilinear
 a simple linear combination. If $ {\bf{q}} \neq 0 $ then,
\begin{equation}
c^{\dagger}_{ {\bf{k}} + {\bf{q}}/2 \sigma }
c_{ {\bf{k}} - {\bf{q}}/2 \sigma^{'} }
 = A_{ {\bf{k}}\sigma }(-{\bf{q}}\sigma^{'}) +
 A^{\dagger}_{ {\bf{k}}\sigma^{'} }({\bf{q}}\sigma)
\end{equation}
\[
 c^{\dagger}_{ {\bf{k}} \sigma }c_{ {\bf{k}} \sigma^{'} }
 = n_{F}({\bf{k}})\delta_{ \sigma, \sigma^{'} }
\]
\begin{equation}
 + \sum_{ {\bf{q}}_{1}\sigma_{1} }
 A^{\dagger}_{ {\bf{k}}-{\bf{q}}_{1}/2 \sigma_{1} }({\bf{q}}_{1}\sigma)
A_{ {\bf{k}}-{\bf{q}}_{1}/2 \sigma_{1} }({\bf{q}}_{1}\sigma^{'})
 -\sum_{ {\bf{q}}_{1}\sigma_{1} }
 A^{\dagger}_{ {\bf{k}}+{\bf{q}}_{1}/2 \sigma^{'} }({\bf{q}}_{1}\sigma_{1})
A_{ {\bf{k}}+{\bf{q}}_{1}/2 \sigma }({\bf{q}}_{1}\sigma_{1})
\label{DIAGPR}
\end{equation}
\noindent
From Eqn.(~\ref{DIAGPR}) it is clear that we must also have,
$ <n^{<}_{ {\bf{k}} \sigma } >  = 1 - S_{B}({\bf{k}}\sigma) $
and $  <n^{>}_{ {\bf{k}} \sigma } > = S_{A}({\bf{k}}\sigma) $.
where,
\begin{equation}
S_{B}({\bf{k}}\sigma) =
\sum_{ {\bf{q}} \sigma_{1} }
\langle G | a^{\dagger}_{ {\bf{k}}+{\bf{q}}/2 \sigma }({\bf{q}}\sigma_{1})
a_{ {\bf{k}}+{\bf{q}}/2 \sigma }({\bf{q}}\sigma_{1}) | G \rangle
(1-n_{F}({\bf{k}}+{\bf{q}}))
\end{equation}
\begin{equation}
S_{A}({\bf{k}}\sigma) =
\sum_{ {\bf{q}} \sigma_{1} }
\langle G | a^{\dagger}_{ {\bf{k}}-{\bf{q}}/2 \sigma_{1} }({\bf{q}}\sigma)
a_{ {\bf{k}}-{\bf{q}}/2 \sigma_{1} }({\bf{q}}\sigma) | G \rangle
n_{F}({\bf{k}}-{\bf{q}})
\end{equation}
 Here $ | G \rangle $ is the ground state of
 the full hamiltonian obtained by treating
 the operators $ A_{ {\bf{k}} \sigma }({\bf{q}}\sigma^{'}) $ more carefully
 and also by retaining the square root of the number operator and
 so on. All this is needed presumably, since
 Luttinger liquids being nonideal, their  momentum distributions fluctuate
 and one must solve the system self-consistently. 
 A simple exercise illustrates this point. Take for example 
 the fluctuation in the number operator
 $ N({\bf{k}},{\bf{k}}) =  <n^{2}_{ {\bf{k}} }> - <n_{ {\bf{k}} }>^2 $.
 From idempotence we may write, 
 $  N({\bf{k}},{\bf{k}}) = <n_{ {\bf{k}} }> ( 1 - <n_{ {\bf{k}} }> ) $.
 For a nonitneracting system at zero temperature this quantity is zero.
 For a Landau Fermi liquid with quasiparticle residue close to unity,
 this quantity is small. However for a Luttinger liquid this quantity
 and is large
 equal to the (square of the) mean itself,
 since in the vicinity of $ |{\bf{k}}| = k_{F} $
 for a Luttinger liquid,  $  <n_{ {\bf{k}}_{F} }> \approx 1/2 $.
 This points to the importance of including fluctuations in the
 momentum distribution. In our work this is done implicitly by 
 resumming the infrared divergences.
 However we have found that the off-diagonal
 counterpart namely,
 $ N({\bf{k}},{\bf{k}}^{'}) =
 <n_{ {\bf{k}} }\mbox{  }n_{ {\bf{k}}^{'} }> -
 <n_{ {\bf{k}} }><n_{ {\bf{k}}^{'} }> $ is vanishingly small
 in the thermodynamic limit.
 A systematic mathematically rigorous
 approach is beyond the scope of this
 article, but the claim is the prescriptions
 in Eq.(~\ref{F1}) and Eq.(~\ref{F2}) capture all these essential features.
 In order to make this plausible, we first note the following feature.
 Since $ 0 \leq S^{0}_{A,B} \leq \infty $, we have
 $ 0 \leq n^{<,>} \leq 1 $, as required for fermions.
 Also in the case of Landau Fermi liquid for weak coupling,
  $ 0 \leq S^{0}_{A,B}({\bf{k}}\sigma)  \ll 1 $, this leads to
 the simple formulas presented above with $ | G \rangle = |G_{0} \rangle $
 and $ S^{0}_{A,B} = S_{A,B} $.
 In the case of a Luttinger liquid,
 the quantities $ S^{0}_{A,B}({\bf{k}}\sigma) $ diverge
 logarithmically in the vicinity of
 $ |{\bf{k}}| = k_{F} $. The use of Eq.(~\ref{F1}) gives us 
 the following momentum distribution of a Luttinger liquid
 in the case of the Luttinger hamiltonian and also in the case of
 the Calogero-Sutherland hamiltonian( here $ \Lambda $ is a momentum cutoff,
 see below ).
\[
{\bar{n}}_{k} = \left[ \frac{1}{2} + \frac{1}{2} \mbox{ }sgn(
k_{F}- |k| )
 \left( \frac{ ||k|-k_{F}|
}{\Lambda} \right)^{\gamma} \right] \theta(\Lambda - ||k|-k_{F}|)
\]
\begin{equation}
 + \mbox{            } \theta(||k|-k_{F}| - \Lambda)\theta(k_{F}-|k|)
\end{equation}
The anomalous exponent $ \gamma $ is characteristic of the model used.
Thus we may now proceed to compute the exponent in either case.

 The other question worth answering is the one posed in the
 beginning, namely is there such a thing
 as an ``Inverse Landau Fermi liquid'' ?
 An examination of Eq.(~\ref{F1}) and Eq.(~\ref{F2})
 tells us that the answer is ``No''.
 This is because we
 find there that
 $ \left< n^{<}_{ {\bf{k}}_{F} \sigma } \right> \geq \left< n^{>}_{ {\bf{k}}_{F} \sigma } \right>  $.
 Thus we have accounted for Landau Fermi liquids, Luttinger liquids (and 
 Wigner crystals) and
 there is no such thing as an Inverse Landau Fermi liquid. 

\section{The Luttinger Model \\ in the Sea-Boson Language}

Here we write down the spinless Luttinger model in the
sea-displacement language. This is then solved by a
straightforward diagonalization
 technique and the momentum distribution is calculated.
The hamiltonian is as follows.
\begin{equation}
H = \sum_{kq} \left( \frac{ k \cdot q }{m} \right)
A^{\dagger}_{k}(q)A_{k}(q)
 + \frac{\lambda}{L} \sum_{q} v(q) \left[ S(q)S(-q) +
 S^{\dagger}(-q)S^{\dagger}(q) \right]
 \label{LUTTHAM}
\end{equation}
 We have tried to retain a close similarity with the notation in Mattis and 
 Lieb\cite{Lieb}.  
Here, the interactions couple only the off-diagonal terms, since
we anticipate that these are responsible for breaking Fermi liquid
behavior. Here $ S(q) = \sum_{k} A_{k}(q) \mbox{     }C(k,q) $
and $ C(k,q) = \theta(\Lambda- |k_{F}-|k+q/2||)
\theta(\Lambda- |k_{F}-|k-q/2||) $.
The cutoff $ \Lambda $ is needed in order to ensure that only
electrons in a shell of
 $ [ \pm k_{F}-\Lambda, \pm k_{F} + \Lambda ] $
  interact and the rest are free.
 We also postulate that $ v(q) = v(0) \theta(2\Lambda - |q| ) $.
 In other words, the delta-function repulsion, but with $ |q| $
 restricted to be small enough. 
 If $ q > 0 $ then we find that the sum over $ k $ in the
 definition of $ S(q) $ is in the region
 $ k \in [k_{F} - \Lambda, k_{F} + \Lambda] $.
 Thus this corresponds to right movers or $ \rho_{1} $ in the
 notation of Lieb and Mattis.
 Similarly, $ S(-q) $ involves the sum peaked at
 $ k = -k_{F} $ and corresponds to left movers. Thus this term couples the
 right-movers to the left-movers and is responsible for breaking
 Fermi-liquid behaviour. 
 In the absence of the $ \theta(2\Lambda-|q|) $ we find due to the
 identity below and the repulsion-attraction duality\cite{Setlur2}
 (more commonly called back scattering)
 the interaction term vanishes identically
 ( $ \sum_{q}S(q)S(-q) \equiv 0 $ ).
 Thus we must really make sure that $ v(k-k^{'}) $ can
 be ignored in comparison with $ v(q) $. This is possible if we include 
 $  \theta(2\Lambda-|q|) $ since then
 $  \theta(2\Lambda-|k-k^{'}|) \approx \theta(2\Lambda-|2k_{F}|) = 0 $.
 We shall not provide the details of the diagonalisation which is by now
 quite well-known to the readers. The dispersion of the 
 collective mode is   $ \omega_{c}(q) = v_{eff} |q| = v_{F} \left( 1 - [ \lambda v(0)/\pi v_{F} ]^{2} \right)^{\frac{1}{2}} |q| $. Thus we arrive
 at the following formula for the sea-boson number operator.
\[
<G_{0}| a^{\dagger}_{k}(q)a_{k}(q) | G_{0} > = 
 \left( \frac{ 2 \pi }{ |q| } \right)
 \left( \frac{ \lambda v(0) }{\pi}  \right)^{2} 
 \frac{1}{L} 
 \mbox{           }
\frac{ C(k,q) }{2 v_{eff} (v_{F} + v_{eff}) } 
\]
\begin{equation}
 =  \left( \frac{2 \pi}{ |q| } \right)
 \frac{1}{2L} 
 \mbox{      }
 C(k,q)
 \left[  \left(1 - \left( \frac{\lambda v(0) }{\pi v_{F}} \right)^{2} \right)
^{-\frac{1}{2}} - 1 \right]
\end{equation}
Thus we see the emergence of the sorts of exponents found by Lieb and Mattis
 \cite{Lieb}. Now we calculate $ S^{0}_{A}(k) \approx S^{0}_{B}(k)
 = S_{0}(k) $ as follows.
\[
S_{0}(k) = \sum_{q} \mbox{   }
<G_{0}| a^{\dagger}_{k-q/2}(q)a_{k-q/2}(q) | G_{0} >
\mbox{        }\theta(k_{F}-|k-q|)
\]
\[
 = \frac{L}{2\pi} \int^{\infty}_{-\infty} dq \mbox{         }
\left( \frac{2 \pi }{ |q| } \right)
 \frac{1}{2L} 
 \mbox{      }
 \theta(\Lambda - |  k_{F} - |k| |)
\theta( \Lambda - | k_{F} - |k-q| |) )
\theta(k_{F}-|k-q|) 
 \left[ \left(1 - \left( \frac{\lambda v(0)}{\pi v_{F}} \right)^{2} \right)
^{-\frac{1}{2}} - 1 \right] 
\]
\begin{equation}
 =  \frac{1}{2} \theta( \Lambda - |k_{F}-|k|| )
 \left[ \left( 1 - \left( \frac{\lambda v(0) }{\pi v_{F}} \right)^{2} \right)
^{-\frac{1}{2}} - 1 \right] 
Ln \left( \frac{\Lambda }{ |k_{F}-|k|| } \right) 
\end{equation}
 The momentum distribution is given by (using Eq.(~\ref{F1}) 
 Eq.(~\ref{F2})  when $ ||k|-k_{F}| < \Lambda $),
\begin{equation}
{\bar{n}}_{k} = \frac{1}{2} \left[ 1 + sgn(k_{F}-|k|) \mbox{        }
\left( \frac{ |k_{F} - |k|| }{\Lambda} \right)^{\gamma}
 \right]
\end{equation}
with the exact exponent completely identical to the one in Mattis and Lieb,
\begin{equation}
\gamma =  \left[ \left( 1 - \left( \frac{\lambda v(0)}{\pi v_{F}} 
\right)^{2} \right)
^{-\frac{1}{2}} - 1 \right] 
\end{equation}
 They\cite{Lieb} set $ v_{F} = 1 $ which we have displayed explicitly here.

\section{ The Calogero Sutherland Model }

 Now we would like to see if other systems can also be Luttinger liquids, like
 the Calogero-Sutherland model (CSM).
 If one uses the the form  $ V_{q} = -\pi\beta(\beta-1)|q|/(2m) $
 (which is the Fourier transform of the inverse square interaction)
 in the formulas in our earlier work,
 we find that the integrals are finite at $ |k| = k_{F} $
 and this means we have a Landau Fermi
      liquid for small $ \beta(\beta-1) $.
      This result contradicts the exact solution via
       Jack polynomials\cite{Calo} which shows
       unequivocally that the CSM is a Luttinger liquid.
       The reason for this fallacy
       is the repulsion attraction duality explained
        in our earlier work\cite{Setlur2}.
 The full hamiltonian consists of a kinetic plus
 exchange part and a correlation part.
 The correlation part in terms of the sea-bosons that takes into account the 
 repulsion attraction duality has already been written down\cite{Setlur2}.
 We try and solve the model when $ 0 < \beta (\beta-1) \ll 1 $.
 In general we would have to diagonalize the nonserparable hamiltonian 
 given below. 
\begin{equation}
H = \sum_{kq} \left( \frac{ k.q }{m} \right)
A^{\dagger}_{k}(q)A_{k}(q) + 
 \sum_{k \neq k^{'} }\sum_{q} \frac{ V_{q} - V_{ k-k^{'} } }{2L}
[A_{k}(-q) + A^{\dagger}_{k}(q)]
 [ A_{k^{'}}(q) + A^{\dagger}_{k^{'}}(-q)]
\label{NONSEP}
\end{equation}        
 In the weak coupling case
  we may write, retaining only the terms that break Fermi
 liquid behaviour,
\begin{equation}
H = \sum_{kq} \left( \frac{ k.q }{m} \right)
A^{\dagger}_{k}(q)A_{k}(q) -\sum_{k \neq k^{'} }\sum_{q} \frac{ V_{2k_{F}} }{2L}
[A_{k}(-q)A_{k^{'}}(q) + A^{\dagger}_{k}(q)A^{\dagger}_{k^{'}}(-q)]
\mbox{      } C(k,q)
\label{CALHAM}
\end{equation}          
 Using ideas given earlier, we may read off the anomalous exponent
 $ \gamma \approx \lambda^{2}v^{2}(0)/2\pi^{2}v^{2}_{F}
 = (\beta(\beta-1))^{2}/8 $.        
 Clearly, for $ 0 < \beta (\beta-1) \ll 1 $ we have $ \gamma \ll 1 $.
 But for $ \beta = 3 $ the anomalous exponent deduced using the Mattis 
 Lieb formula is imaginary. This means we have to include forward
 scattering terms (that is $ V_{q} $ in addition to
 $ V_{k-k^{'}} $)  that are known to make the exponent real no matter
 how strong the coupling is.
 The solution
 of the CSM using Jack polynomials\cite{Calo}
  tells us that $ \gamma < 1 $ for $ \beta = 3 $. 
    \footnote{ 
 This requires quite a bit of effort to show even with the propagator
     written down explicitly, thanks to D. Sen for showing me how. }
      This means we really have to be more careful when
      dealing with systems that have
 $ |\lambda v(0)| > \pi v_{F} $.
 Unfortunately, we have not succeeded in reproducing the results obtained
 via Jack polynomials even qualitatively, apart from being reasonably sure
 that the system is in fact a Luttinger liquid. This is due to the
 non-separable nature of the hamiltonian 
 (Eq.(~\ref{NONSEP})) in the sea-boson language 
 brought about by the need to invoke repulsion-attraction
 duality \cite{Setlur2}.
 We also note that recently Liu\cite{Liu} has shown
 how to compute the asymptotics of the one-particle correlation functions
 in various dimensions for various systems 
 using a powerful method known as eigenfunctional theory. 

\section{ The Luttinger Model in Two Dimensions }

 We solve the analog of the Luttinger model in two space dimensions and
 show that ground state of the system is a Landau Fermi liquid.
 Many authors have in the past addressed this issue. Notable among those
 are works by Anderson\cite{Anderson}, Baskaran\cite{Baskaran} 
 and Bares and Wen\cite{Wen}.
 In the present case, the hamiltonian in the sea-boson language is given by,
\begin{equation}
H = \sum_{ {\bf{k}} {\bf{q}} } \left( \frac{ {\bf{k.q}} }{m} \right)
A^{\dagger}_{ {\bf{k}} }({\bf{q}})A_{ {\bf{k}} }({\bf{q}})
 + \sum_{ {\bf{q}} \neq 0} \frac{ v_{ {\bf{q}} } }{2V}
\sum_{ {\bf{k}} {\bf{k}}^{'} }
[A_{ {\bf{k}} }(-{\bf{q}}) + A^{\dagger}_{ {\bf{k}} }({\bf{q}})]
[A_{ {\bf{k}}^{'} }({\bf{q}}) + A^{\dagger}_{ {\bf{k}}^{'} }(-{\bf{q}})]
\end{equation}
 This hamiltonian describes a self-interacting Fermi gas provided we 
 assume that $ v_{ {\bf{q}} } = \frac{ v_{0} }{m}\mbox{     } \theta(\Lambda - |{\bf{q}}|) $
 where $ \Lambda \ll k_{F} $ here $ 0 < v_{0} \ll 1 $ is a dimensionless 
 parameter. We may solve for the boson occupation number as follows.
\begin{equation}
\left< A^{\dagger}_{ {\bf{k}} }({\bf{q}})A_{ {\bf{k}} }({\bf{q}}) \right>
 = \frac{1}{V}
\sum_{i} \frac{ \Lambda_{ {\bf{k}} }(-{\bf{q}}) }
{ (\omega_{i} + \frac{ {\bf{k.q}} }{m})^{2} }g^{2}_{i}(-{\bf{q}})
\end{equation}
where $ \Lambda_{ {\bf{k}} }(-{\bf{q}}) = n_{F}({\bf{k}}-{\bf{q}}/2)(1 - n_{F}({\bf{k}}+{\bf{q}}/2)) $.
\begin{equation}
g^{-2}_{i}(-{\bf{q}}) = \frac{1}{V} \mbox{    }\sum_{ {\bf{k}} }
 \frac{ n_{F}({\bf{k}}-{\bf{q}}/2) - n_{F}({\bf{k}}+{\bf{q}}/2) }
{ (\omega_{i} - \frac{ {\bf{k.q}} }{m})^{2} }
\end{equation}
\begin{equation}
\epsilon_{RPA}({\bf{q}},\omega) = 
1 + \frac{ v_{ {\bf{q}} } }{V} \sum_{ {\bf{k}} }
 \frac{ n_{F}({\bf{k}}+{\bf{q}}/2) - n_{F}({\bf{k}}-{\bf{q}}/2) }
{ \omega - \frac{ {\bf{k.q}} }{m} }
\end{equation}
 As argued in earlier works, we have to interpret sum over modes with
 care so as to not lose the particle-hole mode, the collective mode
 being obvious. The is particularly important in two dimensions where 
 we expect both to be present. Thus the sum over modes is defined as follows.
\begin{equation}
\sum_{i} \mbox{  }f({\bf{q}},\omega_{i})
  = \frac{ \int^{ \infty }_{0} d \omega \mbox{      }W({\bf{q}},\omega)
 \mbox{      } 
f({\bf{q}},\omega) }
{  \int^{ \infty }_{0} d \omega \mbox{      }W({\bf{q}},\omega) }
\end{equation}
Here the weight function is given by,
\begin{equation}
W({\bf{q}},\omega) = -Im \left( \frac{1}{ \epsilon_{RPA}
({\bf{q}},\omega - i0^{+}) } 
 \right)
\end{equation}
 In our earlier work, we had suggested that in the above formula we
 have to use a dielectric function that is sensitive to
 significant qualitative changes in one-particle properties. 
 The simple RPA dielectric function does not possess qualities
 we expect from a Wigner crystal. Thus we shall have to derive a
 new dielectric function using the localised basis rather than the
 plane-wave basis. In the case of the Fermi gas in two dimensions
 we find that the system is a Landau Fermi liquid and there is no
 need to use a better dielectric function, the simple RPA suffices.
 The momentum distirbution is always given by,
\begin{equation}
< n_{ {\bf{k}} } > = \frac{1}{2} 
 \left[ 1 + e^{ -2S^{0}_{B}({\bf{k}}) } \right] \mbox{          }
n_{F}({\bf{k}}) 
 + \frac{1}{2} 
 \left[ 1 - e^{ -2S^{0}_{A}({\bf{k}}) } \right] \mbox{          }
(1-n_{F}({\bf{k}}))
\label{NBAR}
\end{equation}
\begin{equation}
S^{0}_{A}({\bf{k}}) = \sum_{ {\bf{q}} }
\left< A^{\dagger}_{ {\bf{k}}-{\bf{q}}/2 }({\bf{q}})
A_{ {\bf{k}}-{\bf{q}}/2 }({\bf{q}}) \right>
\end{equation}
\begin{equation}
S^{0}_{B}({\bf{k}}) = \sum_{ {\bf{q}} }
\left< A^{\dagger}_{ {\bf{k}}+{\bf{q}}/2 }({\bf{q}})
A_{ {\bf{k}}+{\bf{q}}/2 }({\bf{q}}) \right>
\end{equation}
The computation of the boson occupation number
 $ \left< A^{\dagger}_{ {\bf{k}} }({\bf{q}})
A_{ {\bf{k}} }({\bf{q}}) \right> $ is the key to evaluating one-particle
 properties. 

 In the case of a Fermi gas in two dimensions with
 short-range repulsive interactions we
 may use the simple RPA dielectric function. 
 The integrals are somewhat complicated
 since in two dimensions, the angular parts are very
 troublesome unlike in three dimensions. 
 Therefore we copy the results first derived by Stern\cite{Stern}.
\begin{equation}
\epsilon^{r}_{RPA}(q,\omega) = 1 + \frac{ m k_{F} v_{q} }{2 \pi q }
\{ \frac{q}{ k_{F} } - C_{-} \left[  \left( \frac{q}{2k_{F}}
 -  \frac{ m \omega }{ k_{F} q }   \right)^2 - 1 \right]^{\frac{1}{2}}
 -  C_{+} \left[  \left( \frac{q}{2k_{F}}
 +  \frac{ m \omega }{ k_{F} q }  \right)^2 - 1 \right]^{\frac{1}{2}} \}
\end{equation}
\begin{equation}
\epsilon^{i}_{RPA}(q,\omega) = \frac{ m k_{F} v_{q} }{ 2 \pi q }
\{ D_{-} \left[ 1 - \left[ \frac{q}{2k_{F}} - \frac{ m \omega }{ k_{F} q }
 \right]^2
 \right]^{\frac{1}{2}}
- D_{+} \left[ 1 - \left[ \frac{q}{2k_{F}} + \frac{ m \omega }{ k_{F} q } 
 \right]^2
 \right]^{\frac{1}{2}} \}
\end{equation}
where,
\begin{equation}
C_{ \pm }  =  sgn \left[ \frac{q}{2k_{F}} \pm \frac{ m \omega }{ k_{F} q } \right],
\mbox{         }
D_{ \pm }  = 0, \mbox{       }
 \left|  \frac{q}{2k_{F}} \pm \frac{ m \omega }{ k_{F} q } \right| > 1 
\end{equation}
\begin{equation}
C_{ \pm }  = 0, \mbox{       } D_{ \pm } = 1, \mbox{       } 
 \left|  \frac{q}{2k_{F}} \pm \frac{ m \omega }{ k_{F} q } \right| < 1 
\end{equation}
\begin{equation}
g^{-2}(-{\bf{q}},\omega) =   \frac{ m k_{F} }{ 2 \pi q } \mbox{     }
 \mbox{     }
\{ \frac{ m C_{-} }{ k_{F} q } 
 \left[  \left( \frac{q}{2k_{F}}
 -  \frac{ m \omega }{ k_{F} q }  \right)^2 - 1 \right]^{-\frac{1}{2}}
 \left( \frac{q}{2k_{F}}
 -  \frac{ m \omega }{ k_{F} q }  \right) 
 -  \frac{ m C_{+} }{ k_{F} q } \left[  \left( \frac{q}{2k_{F}}
 +  \frac{ m \omega }{ k_{F} q } \right)^2 - 1 \right]^{-\frac{1}{2}}
 \left( \frac{q}{2k_{F}} +  \frac{ m \omega }{ k_{F} q }  \right) \}
\end{equation}
\[
\left< A^{\dagger}_{ {\bf{k}} }({\bf{q}})A_{ {\bf{k}} }({\bf{q}}) \right>
 = \frac{1}{V} \frac{1}{ Z(q) }\int^{\infty}_{0} d \omega \mbox{       }W(q,\omega) 
 \mbox{       } \frac{ \Lambda_{ {\bf{k}} }(-{\bf{q}}) }
{ (\omega + \frac{ {\bf{k.q}} }{m})^{2} } \mbox{      } 
 g^{2}(-{\bf{q}},\omega)
\]
\begin{equation}
W(q,\omega) = \frac{ \epsilon^{i}_{RPA}(q,\omega) }
{ \epsilon^{r 2}_{RPA}(q,\omega) 
 +  \epsilon^{i 2}_{RPA}(q,\omega) }
\end{equation}
\begin{equation}
Z(q) = \int^{ \infty }_{0} d \omega \mbox{        }W(q,\omega) 
\end{equation}
In general we have,
\begin{equation}
S^{0}_{A}({\bf{k}}) = \frac{1}{ (2 \pi)^2 } 
\int^{\infty}_{0} dq \mbox{        }q \mbox{     }
\frac{1}{ Z(q) }\int^{\infty}_{0} d \omega \mbox{       }W(q,\omega) 
 \mbox{       } f_{A}(k,q,\omega) 
\mbox{      } 
 g^{2}(-{\bf{q}},\omega)
\end{equation}
\begin{equation}
S^{0}_{B}({\bf{k}}) = \frac{1}{ (2 \pi)^2 } 
\int^{\infty}_{0} dq \mbox{        }q \mbox{     }
\frac{1}{ Z(q) }\int^{\infty}_{0} d \omega \mbox{       }W(q,\omega) 
\mbox{      } 
f_{B}(k,q,\omega)
 g^{2}(-{\bf{q}},\omega)
\end{equation}
 The rest of the details are relegated to Appendix A.
 The collective mode occurs when $ Im[ \epsilon ] = 0 $, that is, for small
 enough $ q $. This means that we have to treat this separately. 
\[
S^{0}_{A}({\bf{k}}) = \frac{1}{ (2 \pi)^2 } 
\int^{\infty}_{0} dq \mbox{        }q \mbox{     }
\frac{1}{ Z(q) }\int^{\infty}_{0} d \omega \mbox{       }W(q,\omega) 
 \mbox{       } f_{A}(k,q,\omega) 
\mbox{      } 
 g^{2}(-{\bf{q}},\omega)
\]
\begin{equation}
 + \frac{1}{ (2 \pi)^2 } 
\int^{\infty}_{0} dq \mbox{        }q \mbox{     } 
 \mbox{       } f_{A}(k,q,\omega_{c}) 
\mbox{      } 
 g^{2}(-{\bf{q}},\omega_{c})
\end{equation}
\[
S^{0}_{B}({\bf{k}}) = \frac{1}{ (2 \pi)^2 } 
\int^{\infty}_{0} dq \mbox{        }q \mbox{     }
\frac{1}{ Z(q) }\int^{\infty}_{0} d \omega \mbox{       }W(q,\omega) 
\mbox{      } 
f_{B}(k,q,\omega)
 g^{2}(-{\bf{q}},\omega)
\]
\begin{equation}
+ \frac{1}{ (2 \pi)^2 } 
\int^{\infty}_{0} dq \mbox{        }q \mbox{     }
f_{B}(k,q,\omega_{c})
 g^{2}(-{\bf{q}},\omega_{c})
\end{equation}
 Here it is implicit that in $ W $ we assume that $ Im[ \epsilon ]  \neq 0 $.
 The dispersion of the collective mode may be found using
$ {\it{ Mathematica }}^{TM} $. It is given below.
\begin{equation}
\omega_{c}(q) = q \mbox{  }(2 \pi + m v_{q}) 
\frac{ (\pi^2 q^2 + m \pi q^2 v_{q} + k_{F}^2 m^2 v_{q}^2)^{\frac{1}{2}} }
{ 2 m^2 \sqrt{ \pi } v_{q} \sqrt{ \pi + m v_{q} } }
\end{equation}
 This dispersion is real and positive for all $ q $ and for all $ v_{q} > 0 $.
 Thus in the small $ q $ limit, where using just the RPA dielectric
 function is justified and is also the limit where
 the close to the Fermi surface features of the momentum distribution is
 given exactly, we are justified in retaining only the coherent
 part. Thus we may write,
\begin{equation}
S^{0}_{A}({\bf{k}}) \approx \frac{1}{ (2 \pi)^2 } 
\int^{\infty}_{0} dq \mbox{        }q \mbox{     } 
 \mbox{       } f_{A}(k,q,\omega_{c}) 
\mbox{      } 
 g^{2}(-{\bf{q}},\omega_{c})
\end{equation}
\begin{equation}
S^{0}_{B}({\bf{k}}) \approx \frac{1}{ (2 \pi)^2 } 
\int^{\infty}_{0} dq \mbox{        }q \mbox{     }
f_{B}(k,q,\omega_{c})
 g^{2}(-{\bf{q}},\omega_{c})
\end{equation}
 To determine whether or not Fermi liquid theory breaks down,
 we have to compute,
\begin{equation}
S^{0}_{A}(k_{F}) \approx \frac{1}{ (2 \pi)^2 } 
\int^{\infty}_{0} dq \mbox{        }q \mbox{     } 
 \mbox{       } f_{A}(k_{F},q,\omega_{c}) 
\mbox{      } 
 g^{2}(-{\bf{q}},\omega_{c})
\label{EQNSAF}
\end{equation}
\begin{equation}
S^{0}_{B}(k_{F}) \approx \frac{1}{ (2 \pi)^2 } 
\int^{\infty}_{0} dq \mbox{        }q \mbox{     }
f_{B}(k_{F},q,\omega_{c})
 g^{2}(-{\bf{q}},\omega_{c})
\label{EQNSBF}
\end{equation}
 If $ S^{0}_{A}(k_{F}), S^{0}_{B}(k_{F}) < \infty $ then the ground state
 is a Landau Fermi liquid.
 If $  S^{0}_{A}(k_{F}) = S^{0}_{B}(k_{F}) = \infty $ then the system
 is a non-Fermi liquid. For small $ q $ if we set
 $ \omega_{c} = v_{eff} \mbox{      }q $ we have,
\begin{equation}
f_{A}(k_{F},q,\omega_{c}) \sim f_{B}(k_{F},q,\omega_{c}) \sim 1/q^2
\end{equation}
Also,
\begin{equation}
g^2(-{\bf{q}},\omega_{c}) \sim q
\end{equation}
 Thus the integrals in Eq.(~\ref{EQNSAF}) and Eq.(~\ref{EQNSBF}) are 
 infrared finite. This means that
 $ S^{0}_{A}(k_{F}), S^{0}_{B}(k_{F}) < \infty $ and the system is a Landau
 Fermi liquid. The details of the momentum distribution can be worked out but
 are not very important. A closed formula for the quasiparticle residue
 may be written down as shown below.
\begin{equation}
Z_{F} = \frac{ e^{ -2 S_{B}^{0}(k_{F}) } +  e^{ -2 S_{A}^{0}(k_{F}) } }{2}
\end{equation}

\section{One Dimensional System with Long-Range Interactions}

 In this section, we consider electrons on a circle interacting via
 a two-body attractive long-range interaction with strength proportional to the
 separation between the electrons. 
 In this case, we expect the system to
 be a Wigner crystal since the interaction is long-range and 
 actually increases (in magnitude) with separation rather
 than decreases. This means
 that the electrons prefer to be as far apart from each other as possible
 to lower the energy leading to a crystalline ground state.
 Thus we have to be careful about
 the choice of the dielectric function.
 First, we postulate that $ v_{ q } = 2e^2/(q\mbox{  }a)^2 $
 which corresponds to the gauge potential. Here $ a $ has
 dimensions of length and $ e^2 > 0 $ is dimensionless. 
 From the form of this potential, one hopes that
 we need not concern ourselves with the issues that
 were relevant in the case of the Calogero-Sutherland model namely the
 repulsion attraction duality. Thus we may write
 as before,
 \[
\left< A^{\dagger}_{ {\bf{k}} }({\bf{q}})A_{ {\bf{k}} }({\bf{q}}) \right>
 = \frac{1}{V} \frac{1}{ Z(q) }\int^{\infty}_{0} d \omega
 \mbox{       }W(q,\omega) 
 \mbox{       } \frac{ \Lambda_{ {\bf{k}} }(-{\bf{q}}) }
{ (\omega + \frac{ {\bf{k.q}} }{m})^{2} } \mbox{      } 
 g^{2}(-{\bf{q}},\omega)
\]
\begin{equation}
W(q,\omega) = \frac{ \epsilon^{i}(q,\omega) }
{ \epsilon^{r 2}(q,\omega) 
 +  \epsilon^{i 2}(q,\omega) }
\end{equation}
\begin{equation}
Z(q) = \int^{ \infty }_{0} d \omega \mbox{        }W(q,\omega) 
\end{equation}
\begin{equation}
 g^{-2}(-{\bf{q}},\omega) = \frac{1}{ v_{q} }
\frac{ \partial }{ \partial \omega } \epsilon({\bf{q}},\omega)
\end{equation}
  As mentioned before,
  we have to be extra careful in making sure that we choose the
 right dielectric function. The RPA-dielectric function is not likely
 to suffice since its static structure factor (SSF) does not exhibit the 
 features we expect from a Wigner crystal. In particular, we expect
 $ S(2k_{F}) = \infty $ as we shall see soon. 
 To convince ourselves of this we ascertain the properties of the 
 RPA dielectric function with long-range interactions.
\begin{equation}
\epsilon^{r}_{RPA}(q,\omega) = 1 + v_{q} \frac{m}{2 \pi q}
Log \left[ \frac{ (k_{F}+q/2)^2 - \left(\frac{m \omega }{q}\right)^2 }
{ (k_{F}-q/2)^2 - \left( \frac{ m \omega }{q} \right)^2 } \right]
\end{equation}
The zero of the above dielectric function gives us the dispersion
 of the collective modes.
\begin{equation}
\omega_{c}(q) = \frac{ |q| }{m} 
\sqrt{ \frac{ (k_{F}+q/2)^2 - (k_{F}-q/2)^2 exp(- \frac{ 2 \pi q }{m v_{q}} ) }{ 1 -  exp(- \frac{ 2 \pi q }{m v_{q} }) } }
\end{equation}
 For $ |q| \ll k_{F} $ and $ v_{q} = 2 e^2/ (a q)^2 $ we find, 
\begin{equation}
\omega_{c}(q) \approx \frac{1}{m} \sqrt{ \frac{ e^2 k_{F} m }{a^2 } }
\sqrt{ \frac{2}{ \pi } }
 + \frac{ a^2 k_{F} \sqrt{ \frac{ e^2 k_{F} m }{ a^2 } }
 \sqrt{ \frac{ \pi }{2} } q^2 }{ 2 e^2 m^2 } + O(q^4)  
\end{equation}
 This plasmon-like gap
 $ \omega_{0} \equiv \frac{1}{m} \sqrt{ \frac{ e^2 k_{F} m }{a^2 } }
\sqrt{ \frac{2}{ \pi } } $ in the collective mode is present due to the
 characteristic $ 1/q^2 $ nature of the potential. But this is also
 present in the three dimensional electron gas and is not a sign of
 an insulator since the latter is not at high densities. A gap in
 the {\it{ one-particle }} Green function could be taken
 as a sign of insulating behaviour\cite{Sen}. However, in our approach we
 are unable to compute the full Green function as yet. Thus we must resort
 to a more indirect approach.
 For a Wigner crystal, the SSF must
 exhibit certain singularities.  
 Thus we have to use the generalised-RPA
 that is sensitive to qualitative changes
 in single-particle properties. The new dielectric function will
 involve the full momentum distribution which has to be determined 
 self-consistently using the above sea-boson equations.
 In our earlier work we suggested that the new dielectric
 function should also
 involve fluctutations in the momentum distribution,
 however it now appears that
 that is fortunately not needed. The number-number
 correlation function is vanishingly small
 in the thermodynamic limit as shown in another preprint
 and this means we may simply write,
\begin{equation}
\epsilon({\bf{q}},\omega) = 1 + \frac{v_{q}}{L}
\sum_{ k } \frac{ {\bar{n}}_{k+q/2} - {\bar{n}}_{k-q/2} }
{ \omega - \xi_{k+q/2} + \xi_{k-q/2} }
\label{WIGDI}
\end{equation}
\begin{equation}
\xi_{k} = \frac{ k^2 }{2m} - \sum_{ q \neq 0 }
 \frac{ v_{q} }{L} {\bar{n}}_{k-q}
\end{equation}
 and the momentum distribution is determined self-consistently
 using the sea-boson equations (Eq.(~\ref{NBAR})).
 This is too difficult to solve analytically and hence we have to resort to
 a numerical solution.
 In order to simplify proceedings even further, we use only the collective
 mode. The particle-hole mode which is due to a nonzero $ Im[\epsilon] $
 is needed if one is interested in features of the momentum distribution away
 from the Fermi surface more accurately. However we shall hope that 
 this is given not too badly even at these regions far from the
 Fermi points. 
\begin{equation}
\left< A^{\dagger}_{ {\bf{k}} }({\bf{q}})A_{ {\bf{k}} }({\bf{q}}) \right>
 = \frac{1}{V} 
 \mbox{       } \frac{ \Lambda_{ {\bf{k}} }(-{\bf{q}}) }
{ (\omega_{c}(q) + \frac{ k.q }{m})^{2} } \mbox{      } 
 g^{2}(-{\bf{q}},\omega_{c})
\end{equation}
\begin{equation}
g^{-2}(-{\bf{q}},\omega_{c}) = \frac{ m \omega_{c} }{ \pi q }
\left[ \frac{1}{ \omega_{c}^2 - (v_{F}q + \epsilon_{q})^2 }
- \frac{1}{ \omega_{c}^2 - (v_{F}q - \epsilon_{q})^2 } \right]
\end{equation}
\begin{equation}
S^{0}_{A}(k) = \frac{1}{L}\sum_{q} 
 \mbox{       } \frac{ n_{F}(k-q) }
{ (\omega_{c}(q) + \frac{ k.q }{m} - \epsilon_{q})^{2} } \mbox{      } 
 g^{2}(-{\bf{q}},\omega_{c})
\end{equation}
\begin{equation}
S^{0}_{B}(k) = \frac{1}{L}\sum_{q} 
 \mbox{       } \frac{ (1-n_{F}(k+q)) }
{ (\omega_{c}(q) + \frac{ k.q }{m} + \epsilon_{q})^{2} } \mbox{      } 
 g^{2}(-{\bf{q}},\omega_{c})
\end{equation}
 To proceed further, we have to ascertain the nature of the collective modes 
 $ \omega_{c} $. If we use the RPA-dielectric function, we find
 a constant dispersion (plasmon) 
 for small $ |q| $. 
 However we have found that this
 choice is inconsistent since if we use the momentum distribution obtained
 from this to solve for the dielectric function and recompute 
 the collective mode we obtain a completely different answer namely :
 $ \omega_{c}(q) = v_{s} |q| $.
 Therefore it is critical that we get the 
 dispersion right. It appears then that we have to use the form given in the
 appendix which is not easy to simplify.
  A systematic approach for obtaining the dispersion
  of the collective modes
  has been suggested by Sen and Baskaran\cite{Baskaran}.
  Since the plane-wave basis is not appropriate for deriving a formula 
  for the dielectric function of a Wigner crystal,
  we shall follow this approach.
  First, we would like ascertain the lattice structure in the small $ a $
  limit. In this limit, the potential energy dominates
  over the kinetic energy. If we assume that the electrons are all on
  a circle of perimeter $ L $ then to minimise the potential energy,
  we have to maximize the separation. This leads to an equally spaced
  set of lattice points with lattice constant $ l_{c} $ such that
  $ N \mbox{       }l_{c} = L $.
  Thus we have $ l_{c} = 1/\rho_{0} = \pi/k_{F} $.
  Thus we assume that the electrons all lie on a circle with equal
  spacing between them. Therefore we expect the structure factor to diverge
 for a momentum $ q_{0} = 2 \pi/l_{c} = 2k_{F} $. From the Bijl-Feynman
 formula $ S(q) = \epsilon_{q}/\omega_{c}(q) $ we may suspect that
 a choice of $ \omega_{c} $ that vanishes at $ q = 2k_{F} $ is 
 needed. The form of the dispersion is given in the appendix.
 For $ \pi/N \ll |q \mbox{   }l_{c}| \ll 2 \pi  $ it seems that
 $ \omega_{q} \approx \omega_{0} $.
 For $ |q \mbox{   }l_{c}| \ll \pi/N  $ we have to be more careful.
 And of course we must have $ \omega_{q} = 0 $ for
 $ q \mbox{   }l_{c} = \pm 2 \pi $.
 But since in the thermodynamic limit $ \pi/N \approx 0 $ we may
 choose (hopefully) $ \omega_{q} \approx \omega_{0} $.
 In Fig. 1 and 2 we see the momentum distribution obtained from these
 formulas has been plotted. In fact, we may write down a closed formula
 for the momentum distribution. 
\begin{equation}
{\bar{n}}_{k} = \frac{1}{2} 
( 1 +  Exp \left[  -\frac{ m \mbox{  }\omega_{0} }
{ k_{F}^2 - k^2 } \right] ) n_{F}(k)
 +  \frac{1}{2} 
( 1 - Exp \left[ -\frac{ m \mbox{  }\omega_{0} }
{ k^2 - k_{F}^2 } \right] ) (1-n_{F}(k))
\end{equation}
 The striking feature of this momentum distribution is that it is perfectly
 flat at $ |k| = k_{F} $. In other words, not only is the slope zero but
 all the derivatives of the momentum distribution vanish at $ |k| = k_{F} $. 
 This is a striking prediction.
 This may be contrasted with the smooth Gaussian function
 of Gori-Giorgi and Ziesche \cite{Gori}( Eq.(B1) in their Appendix B ).
 But they consider three dimensional systems which may be different from
 the one studied here.
 One particle spectral functions
 are accessible to tunneling experiments or angle-resolved
 photoemission spectroscopy(ARPES).
 A more difficult problem may be to
 experimentally realise a 1d electron system with long-range gauge
 interactions. 


\begin{figure}[h]
\centerline{\psfig{file=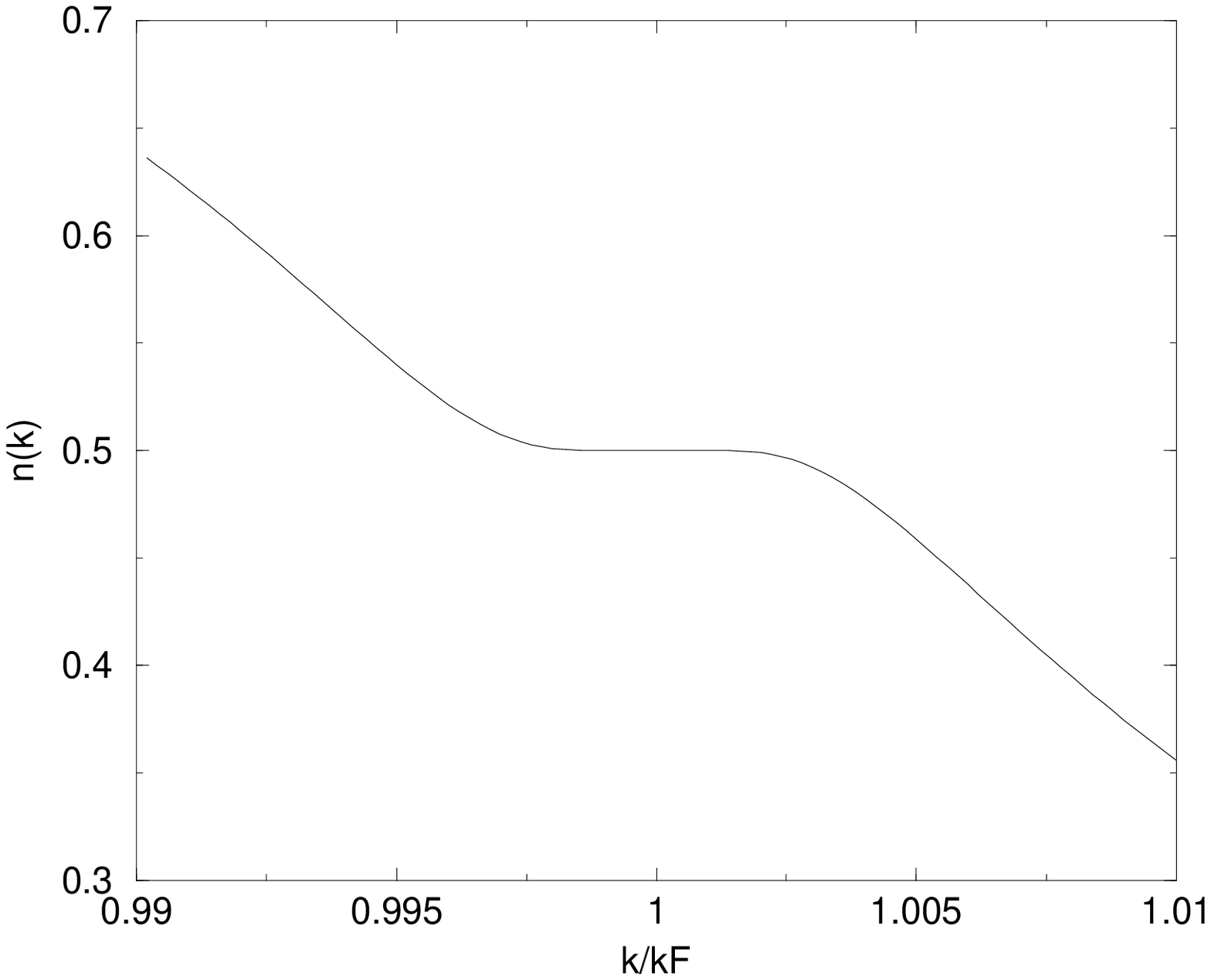,width = 8cm,height=6cm}}
\vspace*{8pt}
\caption { \label{wigner1} Momentum Distribution of a
 Wigner Crystal (with zoom) } 
\end{figure}

\begin{figure}[h]
\centerline{\psfig{file=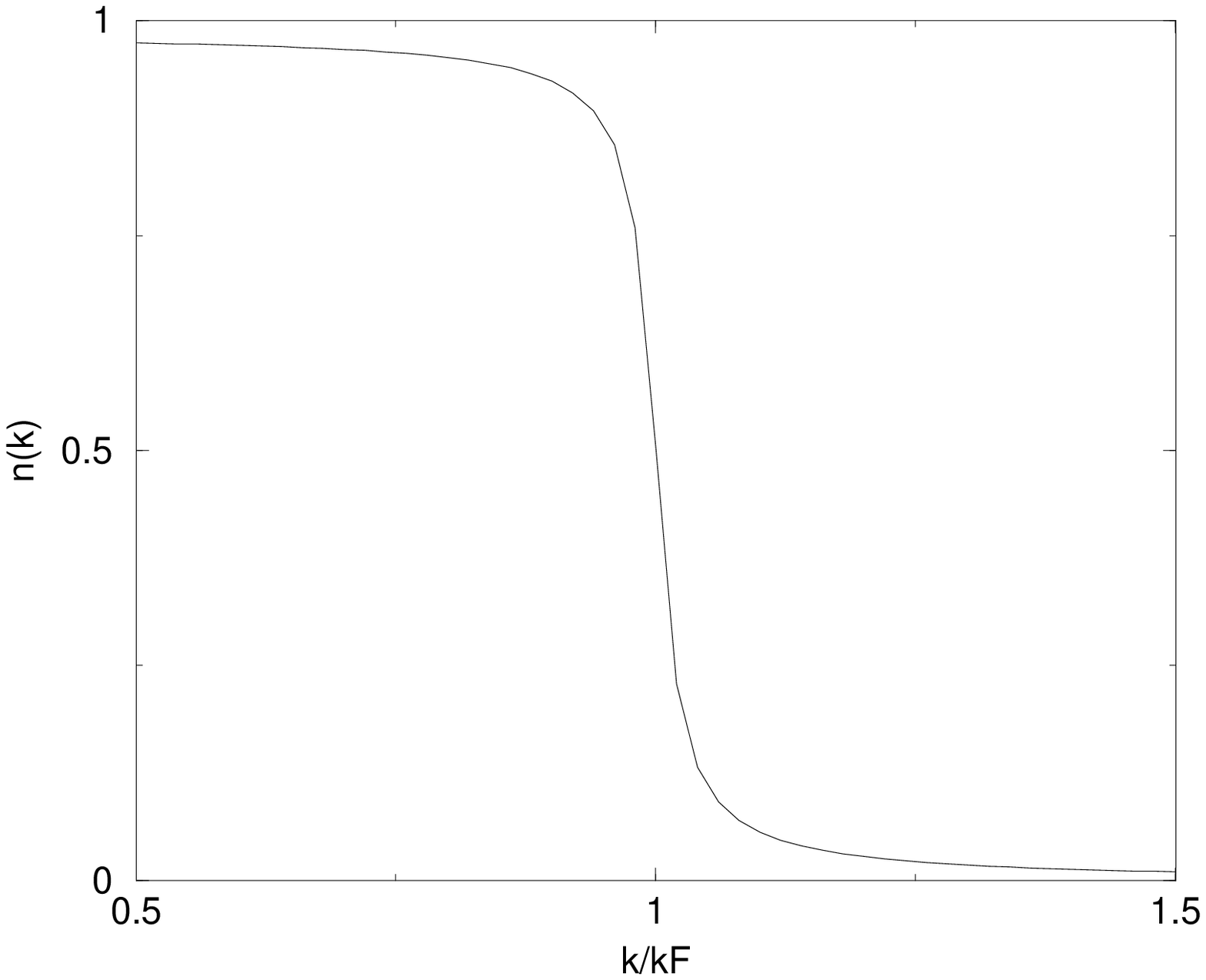,width = 8cm,height=6cm}}
\vspace*{8pt}
\caption{ \label{wigner2} Momentum Distribution of a Wigner Crystal (no zoom) }
\end{figure}
 The formula below for the static structure factor 
  is derived in the appendix.
\begin{equation}
S(q) =  \frac{1}{N} \mbox{   }
 \frac{ sin^2( \frac{ q N \mbox{   }l_{c} }{2} ) }
{  sin^2( \frac{ q \mbox{   }l_{c} }{2} )  }  \mbox{         }
e^{ - \frac{ q^2 }{ m \omega_{0} }  }
\end{equation}
This may be further simplified in the thermodynamic limit as follows.
Consider,
\begin{equation}
\delta(x) \approx \frac{ sin ( N \mbox{   }x ) }{ \pi \mbox{   }x }
\end{equation}
Then we may write,
\begin{equation}
S(q) =   \delta(q) \frac{ 2 \pi }{ l_{c} }
+  \delta(q-2k_{F}) \frac{ 2 \pi }{ l_{c} }
  \mbox{         }
e^{ - \frac{ 4 k_{F}^2 }{ m \omega_{0} }  }
\end{equation}
 As we can see here the strucutre factor
 diverges at $ |q| = 2k_{F} $ which means
 the system is a Wigner crystal ( a true Wigner crystal, since
 the divergence is from a delta-function ). 

\section{Conclusions}

 To conclude, we have finally shown how to reproduce Luttinger liquids
 using sea-bosons. We have also shown how systems with both long-range
 and short-range interactions may be
  studied with almost equal ease and that too
 in any number of dimensions. Using the same formalism we may also 
 probe short-wavelength physics as illustrated by the 
 momentum distribution of the Wigner crystal that contains no
 momentum cutoffs. Finally we showed that the two dimensional
 analog of the Luttinger model is a Landau Fermi liquid.  

\section{Acknowledgements}
 
 The author would like to thank Prof. H.R. Krishnamurthy, Prof. D. Sen
 and Prof. G. Baskaran for critically evaluating the ideas in past works. 
 The author would like to thank Dr. Debanand Sa for sharing his
 extensive knowledge of Many Body Theory and
 for teaching the author how to plot the figures in this article.
 Also Mr. Akbar Jaffari's help with $ Mathematica^{TM} $ 
 is gratefully acknowledged.  

\section{Appendix A}

\begin{equation}
f_{A}(k,q,\omega) = \int^{2 \pi}_{0} d\theta 
\mbox{       }\frac{ \theta( k_{F}^2 - k^2 - q^2 + 2 k q cos(\theta) ) }
{ ( \omega + \frac{ kq }{m} cos(\theta) - \epsilon_{q} )^2 }
\end{equation}
\begin{equation}
f_{B}(k,q,\omega) = \int^{2 \pi}_{0} d\theta 
\mbox{       }\frac{ \theta( k^2 + q^2 -k_{F}^2 + 2 k q cos(\theta) ) }
{ ( \omega + \frac{ kq }{m} cos(\theta) + \epsilon_{q} )^2 }
\end{equation}
Define,
\begin{equation}
u(x;A,B) \equiv \int dx \mbox{       }\frac{ 1 }{ (A + B \mbox{    }Cos[x])^2 }
\end{equation}
From $ Mathematica^{TM} $ we find,
\[
u(x;A,B) = - 2 A \frac{
\mbox{      }ArcTanh \left[ \frac{ (A-B) Tan \left[ \frac{x}{2} \right] }
{ \sqrt{ B^2 - A^2 } } \right] }
{ (A^2 - B^2) \sqrt{ B^2 - A^2 } }
\]
\begin{equation}
 + \frac{ B \mbox{       } Sin[x] } 
{ (B^2 - A^2) (A + B \mbox{   }Cos[x]) }
\end{equation}
Integrating by parts we find,
\begin{equation}
f_{A}(k,q,\omega)
 = \int^{2 \pi}_{0} d\theta \mbox{       }sin \theta \mbox{      }
\mbox{      }\delta( \frac{ k_{F}^2 - k^2 - q^2 }{ 2 k q } + cos(\theta) ) 
\mbox{      }u(\theta ; \omega - \epsilon_{q}, \frac{ k q }{m})
\end{equation}
\begin{equation}
f_{B}(k,q,\omega)
 = \int^{2 \pi}_{0} d\theta  \mbox{       }sin \theta \mbox{      }
\mbox{       }\delta( \frac{ k^2 + q^2 -k_{F}^2 }{ 2 k q } + cos(\theta) ) 
\mbox{       }u(\theta ; \omega + \epsilon_{q}, \frac{ k q }{m})
\end{equation}
This may be rewritten as,
\[
f_{A}(k,q,\omega)
 = \int^{\pi}_{0} d\theta \mbox{       }sin \theta \mbox{      }
\mbox{       } \delta( \frac{ k_{F}^2 - k^2 - q^2 }{ 2 k q } + cos(\theta) ) 
\mbox{     }u(\theta ; \omega - \epsilon_{q}, \frac{ k q }{m})
\]
\begin{equation}
 - \int^{\pi}_{0} d\theta \mbox{       }sin \theta \mbox{      }
\mbox{       } \delta( \frac{ k_{F}^2 - k^2 - q^2 }{ 2 k q } - cos(\theta) ) 
\mbox{     }u(\theta + \pi ; \omega - \epsilon_{q}, \frac{ k q }{m})
\end{equation}
\[
f_{B}(k,q,\omega)
 = \int^{\pi}_{0} d\theta  \mbox{       }sin \theta \mbox{      }
\mbox{       }\delta( \frac{ k^2 + q^2 -k_{F}^2 }{ 2 k q } + cos(\theta) ) 
\mbox{        }u(\theta ; \omega + \epsilon_{q}, \frac{ k q }{m})
\]
\begin{equation}
 - \int^{\pi}_{0} d\theta  \mbox{       }sin \theta \mbox{      }
\mbox{       }\delta( \frac{ k^2 + q^2 -k_{F}^2 }{ 2 k q } - cos(\theta) ) 
\mbox{        }u(\theta + \pi ; \omega + \epsilon_{q}, \frac{ k q }{m})
\end{equation}

\begin{equation}
\theta_{0} = arccos \left[ \frac{ k^2 + q^2 - k_{F}^2 }{2 k q } \right]
\end{equation}
\begin{equation}
\theta^{'}_{0} = arccos \left[ \frac{ - k^2 - q^2 + k_{F}^2 }{2 k q } \right]
\end{equation}
\[
f_{A}(k,q,\omega) = 
 u(\theta_{0} ; \omega - \epsilon_{q}, \frac{ k q }{m})
\left[  \theta( \frac{ k_{F}^2 - k^2 - q^2 }{ 2 k q } + 1 ) 
 -  \theta( \frac{ k_{F}^2 - k^2 - q^2 }{ 2 k q } - 1 )  \right]
\]
\begin{equation}
 - u(\theta^{'}_{0} + \pi ; \omega - \epsilon_{q}, \frac{ k q }{m}) 
 \left[ \theta( 1 - \frac{ k_{F}^2 - k^2 - q^2 }{ 2 k q } ) 
 -  \theta( - 1 - \frac{ k_{F}^2 - k^2 - q^2 }{ 2 k q } )  \right]
\end{equation}
\[
f_{B}(k,q,\omega) = 
 u(\theta^{'}_{0} ; \omega + \epsilon_{q}, \frac{ k q }{m})
\left[ \theta( \frac{ k^2 + q^2 -k_{F}^2 }{ 2 k q } + 1 )
 -  \theta( \frac{ k^2 + q^2 -k_{F}^2 }{ 2 k q } - 1 ) \right]
\]
\begin{equation}
 - u(\theta_{0} + \pi ; \omega + \epsilon_{q}, \frac{ k q }{m})
\mbox{       }
\left[ \theta( 1 - \frac{ k^2 + q^2 -k_{F}^2 }{ 2 k q } ) 
 - \theta( -1 - \frac{ k^2 + q^2 -k_{F}^2 }{ 2 k q } )  \right]
\end{equation}

\section{Appendix B}

 Here we use the approach suggested by Sen\cite{Baskaran} to derive a formula
 for the collective modes.
 The formula Eq.(~\ref{WIGDI}) although probably right is not very
 illuminating, for it is hard to see how the structure factor derived from
 this formula possesses the features we expect namely a divergence
 at $ |q| = 2k_{F} $. Thus we would like to derive a formula for 
 the dielectric function where this feature is manifest. To do this we
 adopt the localised basis rather than the plane-wave basis.
 In real space, the hamiltonian we are studying is written as follows.
\begin{equation}
H = \sum_{i=0}^{N-1} \frac{ p^2_{i} }{2m} - \frac{ e^2 }{ a^2 } \sum_{i > j }
|x_{i}-x_{j}|
\label{HAM}
\end{equation}
 We assume that particles are on a circle and $ |x| $ is the chord length. 
 We would like to compute the dielectric function using this model.
 The density operator in momentum space is,
\begin{equation}
\rho_{ {\bf{q}} } = \sum_{m = 0}^{N-1}e^{i q \left( m \mbox{   }l_{c}
 + {\tilde{x}}_{m} \right) }
\end{equation}
 Here $ x_{m} = m \mbox{   }l_{c} + {\tilde{x}}_{m} $
 is measured along the circumference of the circle
 (see Fig. 3 below).
\begin{figure}[h]
\centerline{\psfig{file=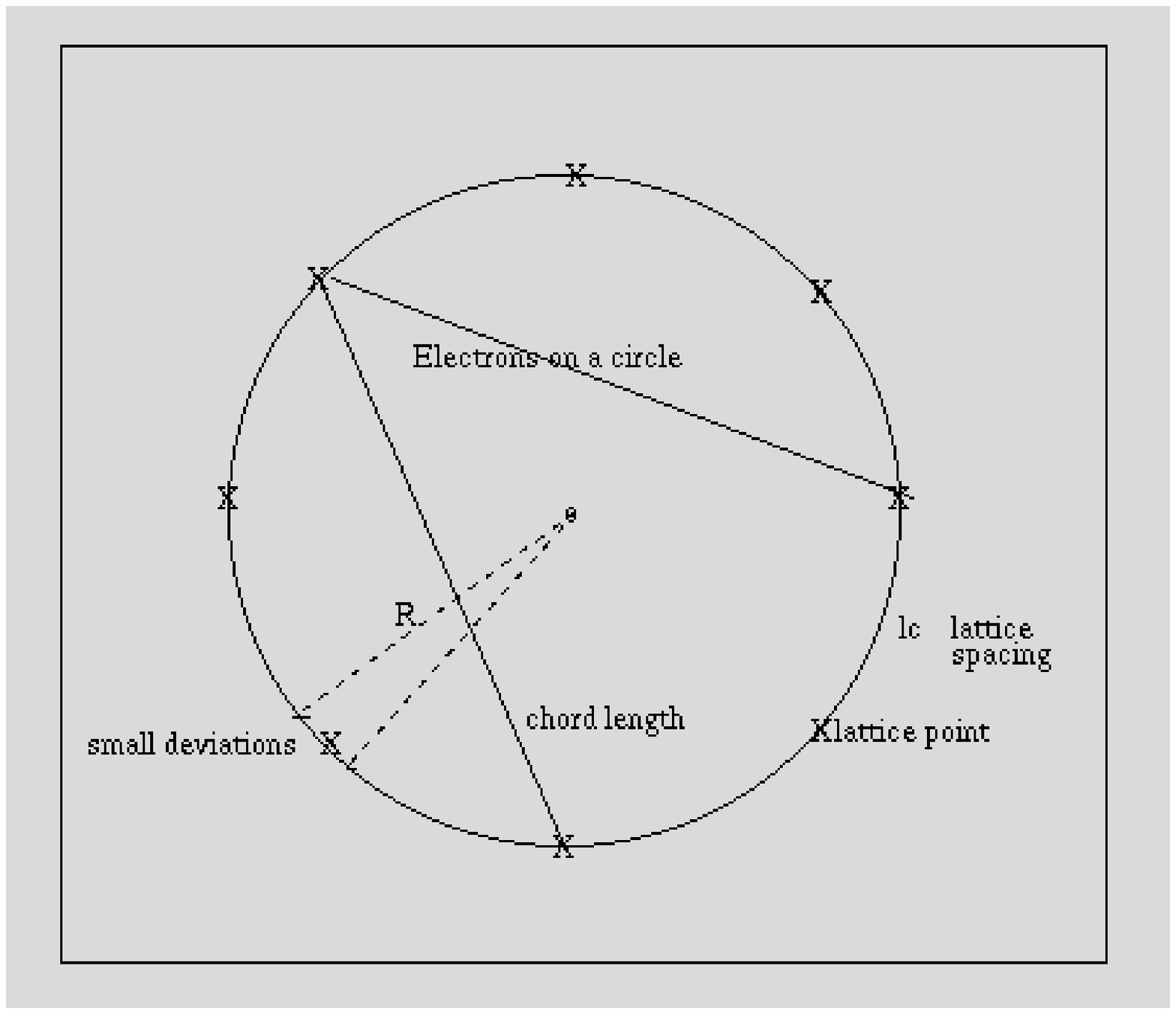,width = 8cm,height=6cm}} 
\vspace*{8pt}
\caption { \label{schema} Schematic Diagram of Electrons on a 
 Circular Lattice }
\end{figure}
 The average density is given by,
\begin{equation}
\left< \rho_{ {\bf{q}} } \right>  
 = \sum_{m = 0}^{N-1}
e^{ i q \mbox{  } m \mbox{   }l_{c}  }
e^{- \frac{1}{2} q^2 \mbox{  }  < {\tilde{x}}^2_{m} > }
\end{equation}
It can be shown that (see below)
 $   < {\tilde{x}}^2_{n} > = 1/m \omega_{0} $, independent of 
the index $ n $.
Thus we have,
\begin{equation}
\left< \rho_{ {\bf{q}} } \right>  
 = \frac{1 - e^{ i q N \mbox{   }l_{c}  } }
{ 1 -  e^{ i q  \mbox{   }l_{c}  } }
e^{ - \frac{ q^2 }{ 2 m \omega_{0} }  }
\end{equation}
 In those instances where $ <\rho_{q}> \neq 0 $ the static structure factor
 is given by,
\begin{equation}
S(q) \equiv \frac{ | <\rho_{ {\bf{q}} }> |^2 }{N} = 
  \frac{1}{N} \mbox{   }
\left( \vline \frac{1 - e^{ i q N \mbox{   }l_{c}  } }
{ 1 -  e^{ i q  \mbox{   }l_{c}  } } \vline \right)^2 \mbox{       }
e^{ - \frac{ q^2 }{ m \omega_{0} }  }
 =  \frac{1}{N} \mbox{   }
 \frac{ sin^2( \frac{ q N \mbox{   }l_{c} }{2} ) }
{  sin^2( \frac{ q \mbox{   }l_{c} }{2} )  }  \mbox{         }
e^{ - \frac{ q^2 }{ m \omega_{0} }  }
\end{equation}
 The rest of the details are as follows. We write
 $ {\tilde{x}}_{m}(t) \approx R \mbox{       }{\tilde{\theta}}_{m}(t) $. 
 In terms of the small angles $ {\tilde{\theta}}_{i} $ the hamiltonian
 in Eq.(~\ref{HAM}) may be written as follows. 
\begin{equation}
H = - \frac{1}{2m R^2} \sum_{m=0}^{N-1} \mbox{   }
\frac{ \partial^2 }{ \partial {\tilde{ \theta }}^2_{m} }
 - \frac{ R \mbox{      }e^2 }{ 2a^2 } \sum_{m \neq m^{'} }
\left[ ( cos ( 2 \pi \frac{ m }{N} + {\tilde{\theta}}_{m} )
 - cos ( 2 \pi \frac{ m^{'} }{N} + {\tilde{\theta}}_{ m^{'} } ) )^2
 + ( sin ( 2 \pi \frac{ m }{N}  + {\tilde{\theta}}_{m}  )
 - sin (2 \pi  \frac{ m^{'} }{N} + {\tilde{\theta}}_{ m^{'} }   ) )^2
 \right]^{\frac{1}{2}}
\end{equation}
 We may expand the above hamiltonian in powers of the angle and retain only
 the leading terms to arrive at the following hamiltonian in the harmonic
 approximation. 
\begin{equation}
 H =  \sum_{n=0}^{N-1} \mbox{   }
\frac{ {\tilde{p}}_{n}^2 }{2m} 
 + \sum_{ n \neq n^{'} }
  A(n,n^{'}) \mbox{         }
( {\tilde{x}}_{n} -  {\tilde{x}}_{ n^{'} })^2
 +  \sum_{ n \neq n^{'} } B(n,n^{'}) \mbox{          }
 ( {\tilde{x}}_{n} -  {\tilde{x}}_{ n^{'} }) 
\end{equation}
where,
\begin{equation}
 A(n,n^{'}) = \frac{ \pi e^2 }{ 4 L a^2 } \mbox{         }
\vline sin( \pi \frac{ (n-n^{'}) }{N} ) \vline
\end{equation}
\begin{equation}
 B(n,n^{'}) = - \frac{ e^2 }{2a^2} 
\mbox{       }sgn( sin( \pi \frac{ (n-n^{'}) }{N} ) )
\mbox{        }cos ( \pi \frac{ (n-n^{'}) }{N} )
\end{equation}
 Despite appearences to the contrary, the extremum of the potential 
 is at $ {\tilde{x}}_{n} \equiv 0 $. Since $ A > 0 $, this extremum is
 also a minimum. One has to now  compute the various correlation
 function of the system. The primary one
 of interest is,
\begin{equation}
 G_{11}(nt;n^{'}t^{'}) = <T  \mbox{  } {\tilde{x}}_{n}(t) \mbox{  } {\tilde{x}}_{n^{'}}(t^{'}) > 
\end{equation}
The other is,
\begin{equation}
 G_{21}(nt;n^{'}t^{'}) =
 <T  \mbox{  } {\tilde{p}}_{n}(t) \mbox{  } {\tilde{x}}_{n^{'}}(t^{'}) > 
\end{equation}
Thus we have,
\begin{equation}
i \frac{ \partial }{ \partial t }  G_{11}(nt;n^{'}t^{'})
 = \frac{i}{m}  G_{21}(nt;n^{'}t^{'})
\end{equation}
\begin{equation}
i \frac{ \partial }{ \partial t }  G_{21}(nt;n^{'}t^{'})
 = \delta_{n, n^{'} } \delta(t-t^{'})
 -4i \sum_{ j \neq n }
A(n,j) ( G_{11}(nt;n^{'}t^{'}) - G_{11}(jt;n^{'}t^{'}) ) 
\end{equation}
This may be solved by a Fourier transform.
\begin{equation}
G_{ij}(nt;n^{'}t^{'}) = \frac{1}{ -i \beta } \sum_{p}
e^{ z_{p} ( t-t^{'} ) } \frac{1}{N} \sum_{q} 
 e^{ i q \mbox{   } l_{c} ( n-n^{'}) } {\tilde{G}}_{ij}(q,z_{p})
\end{equation}
Thus we have,
\begin{equation}
i \mbox{   }z_{p} \mbox{         }  {\tilde{G}}_{11}(q,z_{p})
 = \frac{i}{m}  {\tilde{G}}_{21}(q,z_{p})
\end{equation}
\begin{equation}
i \mbox{   }z_{p}\mbox{   }{\tilde{G}}_{21}(q,z_{p})
 = 1  + 4i \mbox{      }({\tilde{A}}(q)  -{\tilde{A}}(0))
 \mbox{     }{\tilde{G}}_{11}(q,z_{p}) 
\end{equation}
\begin{equation}
{\tilde{A}}(q) = \sum_{j} A(j) e^{ i q \mbox{   }l_{c} \mbox{   }j }
\end{equation}
\begin{equation}
{\tilde{G}}_{11}(q,z_{p}) =
\left( i\mbox{   } m \mbox{   }z_{p}^2  
 -  4i \mbox{      }({\tilde{A}}(q)  -{\tilde{A}}(0)) \right)^{-1}
\end{equation}
Thus,
\begin{equation}
G_{11}(nt;n^{'}t^{'}) = \frac{1}{ N } \sum_{q} 
 e^{ i q \mbox{   } l_{c} ( n-n^{'}) }
\frac{1}{ \beta } \sum_{p}
 e^{ z_{p} ( t-t^{'} ) } 
\left( m \mbox{   }z_{p}^2  
 -  4\mbox{      }({\tilde{A}}(q)  -{\tilde{A}}(0)) \right)^{-1}
\end{equation}
\begin{equation}
{\tilde{A}}(q) = \frac{c_{0}}{2i}
\sum_{j=0}^{N-1} \left( e^{ i ( \frac{ \pi }{N} + q \mbox{   }l_{c})  j }
 -  e^{ i ( -\frac{ \pi }{N} + q \mbox{   }l_{c}) j } \right) 
 = \frac{c_{0}}{2i}
 \left( \frac{ 1 + e^{ i q \mbox{   }l_{c}  N } }
{ 1 - e^{ i ( \frac{ \pi }{N} + q \mbox{   }l_{c}) } }
 -    \frac{ 1 + e^{ i q \mbox{   }l_{c} N  } }
{ 1 - e^{ i ( -\frac{ \pi }{N} + q \mbox{   }l_{c}) } } \right) 
\end{equation}
\begin{equation}
{\tilde{A}}(0) =  \frac{c_{0}}{2i}
 \left( \frac{ 2 }
{ 1 - e^{ i \frac{ \pi }{N} } }
 -    \frac{ 2 }
{ 1 - e^{ -i \frac{ \pi }{N} } } \right)  
 \approx \frac{ k_{F} e^2 }{ 2 \pi  a^2 } 
\end{equation}
 Here $ c_{0} = \pi e^2 / 4 L a^2 $. The dispersion of the collective mode 
 is then given by,
\begin{equation}
\omega_{q} = \sqrt{ \frac{ 4 }{ m } }
\left( {\tilde{A}}(0) - {\tilde{A}}(q) \right)^{\frac{1}{2}}
\label{DISP}
\end{equation}
 If $ q = 0 $ or $ q = 2 k_{F} $ then $ \omega_{q} = 0 $.
 Notice that unless $ |q \mbox{     }l_{c}| \ll \pi/N $, we have
 $ {\tilde{A}}(q) \approx 0 $ for $ |q| \neq 0 $ 
 since it is vanishingly small in the thermodynamic limit. 
 Thus we have to take the large $ N $ limit first in which case
 $ {\tilde{A}}(q) \approx 0 $ for
 $  \pi/N \ll |q \mbox{      }l_{c} | \ll 2 \pi $.
 Thus we may set $ \omega_{q} \approx \omega_{0} $
 with impunity in this case. 
 For a more thorough analysis, one has to compute the full dielectric function
 from the dynamical density-density correlation function and use it to compute the full momentum distribution that
 is accurate even away from the Fermi surface. However we shall be content at
 features close to the Fermi surface.
 For the static structure factor we have to compute the equal time version 
 of the correlation function.  We find that $ \omega_{q} $ is
 in general, complex. This means that the eigenmodes also have a
 finite lifetime. Thus we  have,
\[
G_{11}(nt;n^{'}t) = \frac{1}{ N } \sum_{q} 
 e^{ i q \mbox{   } l_{c} ( n-n^{'}) }
\frac{1}{2 \pi m} \int_{-\infty}^{\infty} d z_{p} \mbox{         } 
\left( z_{p}^2  
 + \omega_{c}^{2}(q) \right)^{-1}
\]
\begin{equation}
 = \frac{1}{ 2m \mbox{  }N } \sum_{q} 
 \frac{ e^{ i q \mbox{   } l_{c} ( n-n^{'}) } }{ \omega_{c}(q) }
\end{equation}
From Eq.(~\ref{DISP}) it is clear that for
 $ \pi/N \ll |q \mbox{  }l_{c}| < 2 \pi $ we have
 $ \omega_{q} \approx \omega_{0} $ since $ {\tilde{A}}(q) \approx 0 $
 for $ q $ in this region. Since in the thermodynamic limit, this
 is all of $ q $, we shall write,
\begin{equation}
< x_{n}(t) \mbox{  }x_{ n^{'} }(t) > 
  = \frac{1}{ 2m \mbox{  }N } \sum_{q} 
 \frac{ e^{ i q \mbox{   } l_{c} ( n-n^{'}) } }{ \omega_{0} }
 = \delta_{ n,n^{'} }
 \mbox{      } 
 \frac{ 1 }{ m \omega_{0} }
\end{equation}

\end{document}